%% file: guettleretal_pg_v01.tex
\documentclass[final]{aipproc}
\layoutstyle{8x11double}
\usepackage{natbib}
\usepackage{amsmath}


\input amssym.def
\input amssym.tex

\begin{document}

\title{Towards a Dynamical Collision Model of Highly Porous Dust Aggregates}


\author{Carsten G{\"u}ttler}{address={Institut f{\"u}r Geophysik und extraterrestrische Physik, TU Braunschweig, Germany}}
\author{Maya Krause}{address={Institut f{\"u}r Geophysik und extraterrestrische Physik, TU Braunschweig, Germany}}
\author{Ralf Geretshauser}{address={Institut f{\"u}r  Astronomie und  Astrophysik, Universit{\"a}t T{\"u}bingen, Germany}}
\author{Roland Speith}{address={Institut f{\"u}r  Astronomie und  Astrophysik, Universit{\"a}t T{\"u}bingen, Germany}}
\author{J{\"u}rgen Blum}{address={Institut f{\"u}r Geophysik und extraterrestrische Physik, TU Braunschweig, Germany}}

\begin{abstract}
In the recent years we have performed various experiments on the
collision dynamics of highly porous dust aggregates and although
we now have a comprehensive picture of the micromechanics of those
aggregates, the macroscopic understanding is still lacking. We are
therefore developing a mechanical model to describe dust aggregate
collisions with macroscopic parameters like tensile strength,
compressive strength and shear strength. For one well defined dust
sample material, the tensile and compressive strength were
measured in a static experiment and implemented in a Smoothed
Particle Hydrodynamics (SPH) code. A laboratory experiment was
designed to compare the laboratory results with the results of the
SPH simulation. In this experiment, a mm-sized glass bead is
dropped into a cm-sized dust aggregate with the previously
measured strength parameters. We determine the deceleration of the
glass bead by high-speed imaging and the compression of the dust
aggregate by x-ray micro-tomography. The measured penetration
depth, stopping time and compaction under the glass bead are
utilized to calibrate and test the SPH code. We find that the
statically measured compressive strength curve is only applicable
if we adjust it to the dynamic situation with a ``softness''
parameter. After determining this parameter, the SPH code is
capable of reproducing experimental results, which have not been
used for the calibration before.
\end{abstract}

\maketitle

\section{Introduction}
Planets form from micrometer-sized dust grains, colliding at low
velocities and sticking to each other due to attracting
van-der-Waals forces \cite{BlumWurm:2008}. From the interplay
between laboratory experiments \cite{BlumWurm:2000} and molecular
dynamic simulations \cite{PaszunDominik:2008,DominikTielens:1997}
we have a good picture of the microphysics of those processes, but
as aggregates grow larger (e.g. 1~mm), the understanding of the
collision mechanics is severely lacking. For these sizes (and the
corresponding velocities), collisions between equal sized
aggregates rather lead to bouncing or fragmentation than to
sticking.

Millimeter-sized aggregates -- as precursors of planets -- are
expected to be highly porous, having a volume filling factor (the
volume fraction of material) of only few percent up to few ten
percent. This strongly cohesive material is comparable to
millimeter-sized dust clumps present in conventional dry powders
(e.g. cocoa, confectioners' sugar) but probably being more porous.
One well analyzed analog material are dust aggregates formed by
the random ballistic deposition (RBD) method, experimentally
introduced in Ref. \cite{BlumSchraepler:2004}. These dust
aggregates have a volume filling factor of $\phi=0.15$ (for
monospheric dust) and are produced in our laboratory in
macroscopic 2.5~cm samples. Using these dust samples, we performed
various collision experiments to study the further growth and
evolution of planetary bodies
\cite[e.g.][]{BlumWurm:2008,LangkowskiEtal:2008}. However, as
experiments cannot be performed for any relevant set of parameters
(e.g. collisions of meter-sized bodies), a collision model is
required to cover the wide parameter space occurring for
protoplanetary dust-aggregate collisions, i.e. dust aggregate
sizes of up to 1~km and collision velocities in the range of
$10^{-3} - 10^2\;{\rm m\;s^{-1}}$. The approach is therefore to
measure macroscopic parameters of the laboratory dust samples and
implement them in a numeric simulation of a dust aggregate
collision, which is concurrently performed in a laboratory
experiment. Comparing experiment and simulation and defining and
determining the free parameters, yields a calibrated collision
model to perform collision simulations with parameters
unaccessible to laboratory experiments.

\section{Smoothed Particle Hydrodynamics for Dust Collisions}
For the simulation of dust aggregate collisions we use the
Smoothed Particle Hydrodynamics (SPH) method with extensions for
the treatment of solid and porous media. A comprehensive
description of the meshless Lagrangian particle method SPH can for
example be found in \cite{monaghan:2005}. In this scheme the
continuous solid objects are discretized into interacting mass
packages (``particles'') carrying all relevant continuous
quantities. Time evolution is computed according to the Lagrangian
equations of continuum mechanics:
\begin{eqnarray}
    \frac{\mathrm{d}\varrho}{\mathrm{d}t} + \varrho \sum_{\alpha =
    1}^{D} \frac{\partial v_\alpha}{\partial x_\alpha} = 0\\
    \frac{\mathrm{d} v_\alpha}{\mathrm{d}t} = \frac{1}{\varrho}
    \sum_{\beta = 1}^{D} \frac{\partial \sigma_{\alpha\beta}}{\partial
    x_\beta}
\end{eqnarray}
Here, $\varrho$ denotes the density, $v$ the velocity, $D$ the
dimension and $\sigma_{\alpha\beta}$ the stress tensor, defined as
\begin{equation}
    \sigma_{\alpha\beta} = -p \delta_{\alpha\beta} + S_{\alpha\beta}.
\end{equation}
It consists of a pressure part with pressure $p$ and and a shear
part given by the traceless deviatoric stress tensor
$S_{\alpha\beta}$. Its time evolution is modelled according to
Ref. \cite{benz:1994}. This set of equations is closed by a
suitable equation of state and describes the elastic behavior of a
solid body. Together with a suitable damage model, which we do not
adopt, the authors of Ref. \cite{benz:1994} have modelled
collisions between brittle basaltic rocks using this scheme.

In contrast to this, we simulate the plastic behavior of porous
bodies. Therefore we adopt a modified version of the porosity
model by Sirono \cite{Sirono:2004}. According to this approach,
plasticity is modelled within the equation of state and porosity
is given by $1 - \varrho/\varrho_0$, where $\varrho$ denotes the
actual and $\varrho_0$ the bulk density of the material. The
pressure is limited by the compressive strength $\Sigma(\varrho)$
as upper bound and the (negative) tensile strength $T(\varrho)$ as
lower bound. In between, the solid body experiences elastic
deformation, whereas outside this regime it is deformed
plastically. Thus, the full equation of state reads
\begin{equation}
    p(\varrho) = \left\{
    \begin{array}{lc}
    \Sigma(\varrho) & \varrho > \varrho_{\rm c}^+ \\
    K({\varrho'}_0) (\varrho/{\varrho'}_0 - 1) & \varrho_{\rm c}^- \le \varrho \le \varrho_{\rm c}^+ \\
    T(\varrho) & \varrho < \varrho_{\rm c}^- \\
    \end{array}
    \right.
\end{equation}
The quantity ${\varrho'}_0$ denotes the density of the material at
zero external stress. $\rho_{\rm c}^+$ and $\rho_{\rm c}^-$ are
limiting quantities, where the transition between the elastic and
plastic regime for compression and tension, respectively, takes
place. Once a limit is exceeded, the material leaves the elastic
path where energy is conserved, and loses internal energy by
following the paths of the compressive and tensile strength.

In a previous work, Sirono \cite{Sirono:2004} adopted power laws
from measurements with toner particles for $\Sigma(\varrho)$,
$T(\varrho)$ and the bulk modulus $K(\varrho)$ in order to
simulate porous ice, which is a crude extrapolation. For our
approach, we used the material properties, measured for well
defined RBD dust aggregates \cite{BlumSchraepler:2004}.
\begin{figure}
    \includegraphics[height=\columnwidth,angle=90]{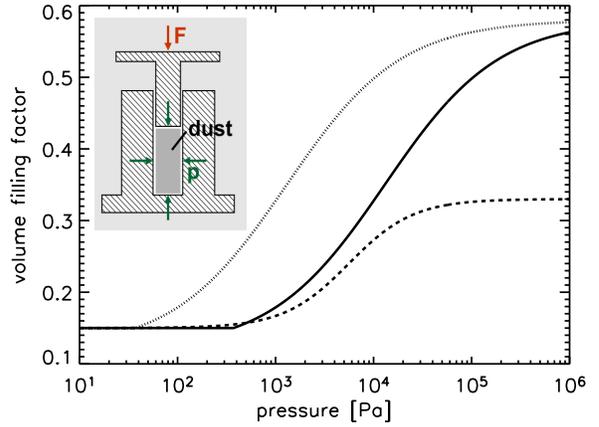}
    \caption{Compressive strength curve for unidirectional (dashed
    line), omnidirectional (solid line), and dynamic (dotted line)
    compression. The inset illustrates the setup for the
    omnidirectional compression measurement.}
    \label{fig-compressive_strength}
\end{figure}
For the tensile strength we adopted the measurement of Ref.
\cite{BlumSchraepler:2004} who measured the tensile strength for
porous ($\phi=0.15$) and compact ($\phi=0.54$) aggregates and
found an agreement with a linear dependence between tensile
strength and numbers of contact per cross-sectional area. This
yields a tensile strength of
\begin{equation}
    T(\phi) = - \left( 10^{2.8 + 1.48 \phi}\right)\;{\rm Pa.}
    \label{eq-tensile_strength}
\end{equation}
For the compressive strength we started with the compression curve
$\phi(\Sigma)$ measured in Ref. \cite{BlumSchraepler:2004}
(unidirectional, 1D compression) and also made a new compression
measurement (omnidirectional, 3D, see inset in Fig.
\ref{fig-compressive_strength}, \cite{GuettlerEtal:inprep}). Both
compressive strength curves are displayed in Fig.
\ref{fig-compressive_strength}. They resemble in shape and can be
described by
\begin{equation}
    \Sigma(\phi)=p_{\rm m}\cdot
    \left(\frac{\phi_2-\phi_1}{\phi_2-\phi}-1\right)^{\Delta\cdot\ln
    10}\;,\label{eq-compression_curve}
\end{equation}
with four free parameters $\phi_1$, $\phi_2$, $\Delta$, and
$p_{\rm m}$. For unidirectional compression we found $p_{\rm
m}=5.6$~kPa, $\phi_1=0.15$, $\phi_2=0.33$, and $\Delta=0.33$,
while for ominidirectional compression measurements performed in
this work the parameters are identified as $p_{\rm m}=13$~kPa,
$\phi_1=0.12$, $\phi_2=0.58$, and $\Delta=0.58$. For
unidirectional, static compression, the material can creep
sideways, releasing pressure. As we did not observe this in the
dynamic experiments (next section, Fig.
\ref{fig-density_contour}), we expect omnidirectional compression,
defining the parameters $\phi_1$ and $\phi_2$. For low pressures,
$\Delta\cdot\ln10$ is the slope of a power law found by divers
authors \cite{BlumSchraepler:2004,ValverdeEtal:2004} and does not
leave very much margin. Thus, we take $p_{\rm m}$ as the only free
parameter with the most influence on material softness, shifting
the compression curve towards lower pressures and determine
$p_{\rm m}$ within the calibration procedure. As the the shear
strength was not measured so far, we follow Sirono
\cite{Sirono:2004} and take $Y=\sqrt{\Sigma|T|}$.

\section{Calibration Experiment}
\begin{figure}
    \includegraphics[height=\columnwidth,angle=90]{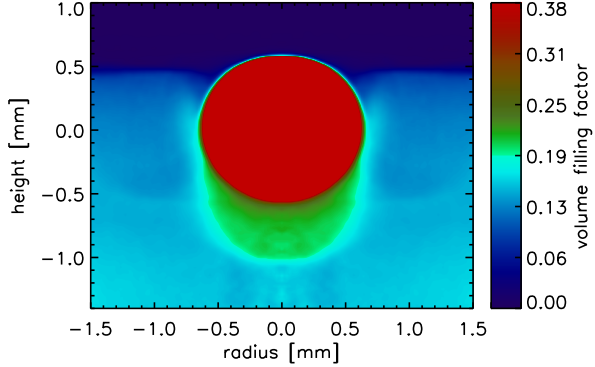}
    \caption{The density plot reveals the compaction of the dust
    sample under the glass bead (red, saturated) with a volume filling
    factor of 0.20 to 0.25 (green) in a well confined volume of
    approximately one sphere volume. The original dust material has a
    volume filling factor of 0.15 (light blue).}
    \label{fig-density_contour}
\end{figure}
In the experiments, a solid projectile was dropped into a 2.5~cm
diameter, highly porous dust aggregate consisting of $1.5\;\mu$m
SiO$_2$ monomers (RBD dust aggregate from
\cite{BlumSchraepler:2004}). The experiments were performed in
vacuum ($0.1$~mbar) such that gas effects did not play a role. The
projectile was either a glass bead of 1~mm diameter or a
cylindrical plastic tube with a 1 or 3~mm diameter epoxy droplet
at the bottom (representing a 1 or 3~mm glass bead). The epoxy
projectile had a mass corresponding to a glass bead and therefore
was longer and could be observed for a penetration deeper than its
diameter. For 15 experiments with elongated projectiles, the
deceleration curve and, thus, the penetration depth and the
stopping time were measured with a high-speed camera. The stopping
time for one projectile diameter was found to be rather constant,
($3.0\pm0.1$)~ms for 1~mm projectiles and ($6.2\pm0.1$)~ms for
3~mm projectiles, while the penetration depth depended on the
projectile size and the impact velocity (drop height). Details on
the full deceleration curve can be found in
\cite{GuettlerEtal:inprep}. The penetration depth could well be
approximated by
\begin{equation}
    D=\left(8\cdot10^{-4}\;\frac{\rm m^2\;s}{\rm kg}\right)\cdot\frac{mv}{A}\;,\label{eq-penetration_depth}
\end{equation}
where $v$, $m$, and $A=\pi R^2$ are the projectile velocity at the
time of first contact, the projectile mass, and its maximum
cross-sectional area (see also Fig. \ref{fig-penetration_depth} in
the next section). The stopping time and the relation for the
penetration depth will be used for the calibration later on.

Two experiments with a 1~mm glass bead impacting with
$(0.8\pm0.1)$~m~s$^{-1}$ were analyzed using an x-ray
micro-tomograph (Micro-CT SkyScan 1074). In this method, the dust
sample with the embedded glass bead was positioned between an
x-ray source and a detector and rotated stepwise. Based on the 400
resulting transmission images, a 3-dimensional density
reconstruction was calculated using the SkyScan Cone-Beam
Reconstruction Software. Assuming cylindrical symmetry in the axis
of penetration, the density was averaged to one vertical section
and displayed in the contour plot in Figure
\ref{fig-density_contour}. The compaction of the dust is clearly
visible in a confined volume under the sphere. The green color
marking this volume denotes a volume filling factor of 0.20 to
0.25, while the material around this compaction zone is virtually
unaffected with an original volume filling factor of
$\phi\approx0.15$. The distribution of compacted material will be
used for the verification of the SPH code.

\section{Calibration and Verification of the SPH Code}
In the 2D SPH simulation, an infinite cylinder with 1.1~mm
diameter ($\rho=2540$~kg~m$^{-3}$) impacts into a dust sample with
a cross section of $8\times5\;{\rm mm}^2$ at a velocity of
0.65~m~s$^{-1}$. For comparison with the 3D experiments a
correction factor of $\frac{8}{3\pi}$ for the penetration depth is
required (see \cite{GuettlerEtal:inprep}). For the calibration we
will use the penetration depth, which has to be 0.82~mm (Eq.
\ref{eq-penetration_depth} and correction factor), the stopping
time of 3~ms, and the compaction under the glass bead from the
x-ray experiments.

\begin{figure}
    \includegraphics[height=\columnwidth,angle=90]{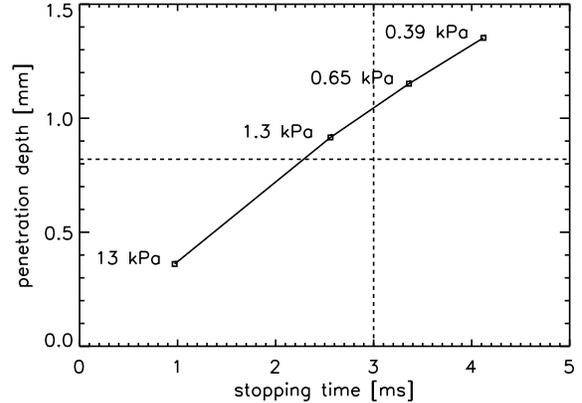}
    \caption{Varying the parameter $p_{\rm m}$ in the simulation
    changes the softness of the material and yields different
    penetration depths and stopping times. The best agreement with the
    experiments was found for $p_{\rm m}=1.3$~kPa.}
    \label{fig-time_intrusion_calibration}
\end{figure}

Using the shear strength formalism introduced by Sirono
\cite{Sirono:2004}, e.g. $Y=\sqrt{\Sigma|T|}$, we have one free
parameter with influence on the outcome of the simulations, i.e.
the adjustment of the compressive strength curve for dynamic
compression by the parameter $p_{\rm m}$ (see Eq.
\ref{eq-compression_curve}). A detailed study on the shear model
can be found in \cite{GuettlerEtal:inprep}.

The parameter $p_{\rm m}$ defines the softness of the material and
decreasing this parameter yields deeper penetrations. Figure
\ref{fig-time_intrusion_calibration} describes the calibration of
this parameter. The horizontal line denotes the expected
penetration depth from Eq. \ref{eq-penetration_depth}, while the
vertical line represents a mean stopping time of 3~ms for 1~mm
spheres. A value of $p_{\rm m}=1.3$~kPa yields the best agreement
with the experiments and thus we will use this value for further
tests.

One test for the validation of the SPH code is the penetration
depth relation for different sizes and different velocities,
reproducing Eq. \ref{eq-penetration_depth}.
\begin{figure}
    \includegraphics[height=\columnwidth,angle=90]{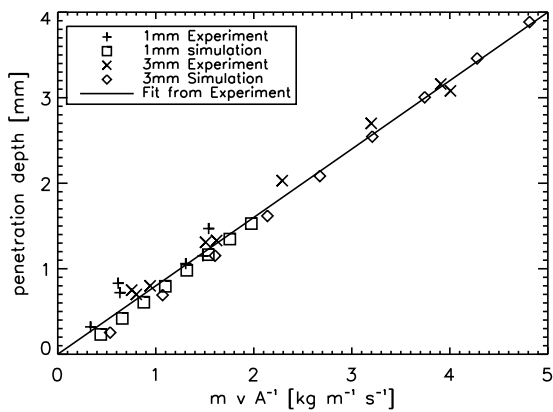}
    \caption{The penetration depth as a function of momentum per
    cross-sectional area can well be reproduced with the SPH code.}
    \label{fig-penetration_depth}
\end{figure}
The penetration depths of the experiments and the penetration depths of the simulations (with correction factor) are plotted in Fig.
\ref{fig-penetration_depth}. For $mvA^{-1}\gtrsim 1$~kg~m$^{-1}$~s$^{-1}$ experiment and simulation are in very good agreement and the SPH code
succeeds in the scaling of radius and velocity. A second validation test is the comparison of the compressed volume with the results of the x-ray
measurement (Fig. \ref{fig-density_contour}).
\begin{figure}
    \includegraphics[height=\columnwidth,angle=90]{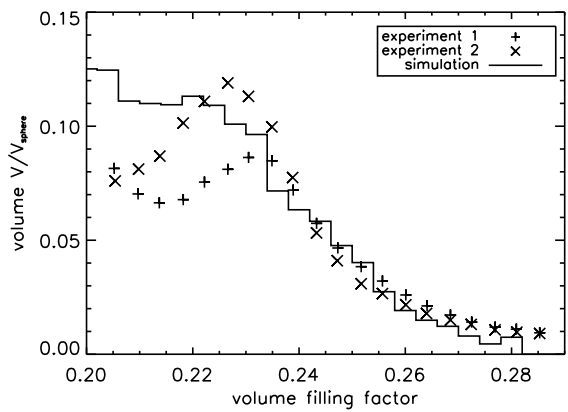}
    \caption{Experimental results (symbols) and simulation (solid
    line) of the compressed volume.}
    \label{fig-compr_volume}
\end{figure}
From the 3-dimensional density data, one can determine the volume
fraction that is compressed to a given volume filling factor,
which is plotted in Fig. \ref{fig-compr_volume}. The values for
$\phi\lesssim0.2$ represent the original material being
distributed around the mean filling factor of 0.15. Also for this
test, we find a good agreement between experiment and simulation.

\section{Conclusion}
We have developed an SPH code for the description of collisional
interaction between high-porosity dust aggregates. The code was
calibrated by static and dynamic laboratory experiments using
macroscopic RBD aggregates with an uncompressed filling factor of
$\phi=0.15$. The calibrated SPH code correctly predicts the size
and velocity dependence of the penetration depth for an impacting
solid projectile as well as the compressed volume.

\begin{theacknowledgments}
We thank M.-B. Kallenrode and the University of Osnabr{\"u}ck for
providing access to the XRT setup. Simulations were performed on
clusters of the computing center (ZDV) of the University of
T{\"u}bingen. This project is funded by the Deutsche
Forschungsgemeinschaft within the Forschergruppe 759 ``The
Formation of Planets: The Critical First Growth Phase'' under
grants Bl 298/7-1, Bl 298/8-1, and Kl 650/8-1.
\end{theacknowledgments}

\bibliographystyle{aipproc}
\bibliography{literatur}

\end{document}